\documentclass[conference]{IEEEtran}
\usepackage{cite}
\usepackage{amsmath,amssymb,amsfonts}
\usepackage{algorithmic}
\usepackage{graphicx}
\usepackage{textcomp}
\usepackage{xcolor}
\usepackage{booktabs}
\usepackage{multirow}
\usepackage{tikz}
\usepackage{tipa}
\usetikzlibrary{positioning, arrows.meta, fit, shapes.geometric, calc}
\usepackage{url}

\def\BibTeX{{\rm B\kern-.05em{\sc i\kern-.025em b}\kern-.08em
    T\kern-.1667em\lower.7ex\hbox{E}\kern-.125emX}}
\begin{document}

\title{ART-NAD: An Articulatory Inversion-based Neural Acoustic Distance for Pathological Speech Intelligibility Assessment}

\author{\IEEEauthorblockN{Bence Mark Halpern}
\IEEEauthorblockA{\textit{Nagoya University}\\
Nagoya, Japan}
\and
\IEEEauthorblockN{Thomas Tienkamp}
\IEEEauthorblockA{\textit{University of Cologne}\\
Cologne, Germany}
\and
\IEEEauthorblockN{Defne Abur}
\IEEEauthorblockA{\textit{University of Groningen}\\
Groningen, the Netherlands}
\and
\IEEEauthorblockN{Tomoki Toda}
\IEEEauthorblockA{\textit{Nagoya University}\\
Nagoya, Japan}}

\maketitle

\begin{abstract}
Speech assessment tools for speakers with speech pathology must be both accurate and interpretable if they are to be adopted in clinical practice. Existing reference-audio measures such as the Neural Acoustic Distance (NAD) reach high speaker-level correlations with listener intelligibility scores but operate on self-supervised features that are hard to interpret, providing only frame-level explanations. We propose ART-NAD, a reference-audio intelligibility metric that replaces the \texttt{wav2vec2} features of NAD with vocal-tract constriction variables (tract variables, TVs) predicted from audio by a speaker-independent acoustic-to-articulatory inversion model trained on the same \texttt{wav2vec2} features. ART-NAD is computed as the multivariate Dynamic Time Warping distance between the nine-channel quasi-TV trajectories of the test and one or more references. Across 20 reference-audio protocols spanning six pathological-speech datasets and five languages, ART-NAD with silence trimming (ART-NAD-FA) reaches the same average speaker-level Pearson correlation as NAD-FA (both $r=0.71$) on the same self-supervised backbone, with no significant per-protocol difference (Wilcoxon $p=0.18$), and is the strongest reference-audio metric on 6 of the 20 protocols. Beside the score itself, each TV channel visualizes which constriction deviates from the reference over time, providing interpretable information as to where articulation breaks down.
\end{abstract}


\section{Introduction}

Assessing speech intelligibility of individuals with speech disorders is important to monitor speech function \cite{orozco2020apkinson}, and to assess effectiveness of interventions \cite{mendoza2021effect}. Currently subjective methods are not supplemented by objective assessments in clinical practice, even though interrater reliability of subjective methods is often low and influenced by listener exposure \cite{landa2014association} and expertise \cite{de1997test}.
Despite the existence of objective methods, there is still a lack of clinical professionals using automated tools for speech evaluation, which is largely due to a lack of transparency in AI evaluation models \cite{gurevich2017speech, liss2024operationalizing}.

Several attempts have been made to improve interpretability, so that a clinician can see why a recording received a particular score. The most common approach uses automatic speech recognition (ASR): the system transcribes the recording, and intelligibility is reported as the fraction of phonemes (or words) that the system recognized correctly \cite{yeo2026multilingual}. A second approach, the Neural Acoustic Distance (NAD), compares the recording against a typical reference recording of the same sentence and produces a score based on how far apart their features are; it also points to the time segments where the two recordings diverge most strongly \cite{bartelds2022neural}.

Another approach is to estimate articulatory-acoustic properties from the speech signal, which has interested the speech community since the source-filter model \cite{fant1971acoustic}. For example, vowel formants have been related to tongue and jaw positioning, with $F_1$ tracking how open the vocal tract is and $F_2$ how far forward the tongue is. As such, the spread of a speaker's vowel formants, summarized as the vowel space area (VSA), is widely used to assess articulation \cite{vanson18_interspeech, ferguson2007talker}. Formants are only an indirect measure of the articulators, though, and are difficult to estimate as they degrade on noisy speech quickly. Good formant estimation is often only achievable by manually tuning formant ceilings, which is not ideal in cases where automatic solutions are needed.

A parallel line of work predicts phonological features, such as manner and place of articulation, from audio with deep neural networks, and these features have been shown to correlate strongly with subjective intelligibility ratings \cite{chen22l_interspeech, middag2011combining}. While these features are informative about global articulation problems, they
cannot localize the actual breakdown in articulation: they are categorical
phonological labels associated with phoneme labels rather than measurements of
the vocal tract. Thus, an erroneous feature indicates which phonological contrast
was not realized, but not which articulator deviated or by how much.

Fortunately, there are other methods to estimate articulatory properties directly. Electromagnetic articulography (EMA) tracks the position of small sensors glued to the articulators (e.g.\ tongue, lips) during speech, capturing the movement of the vocal tract; EMA has been widely used in articulatory phonetics, both in speakers with and without speech disorders \cite{tienkamp2025associations, rebernik2021review}. EMA features have also been used to predict intelligibility in ALS with support vector machine classifiers \cite{wang2018automatic}, and EMA estimated from audio (quasi-EMA) has been used to estimate Parkinson's severity with support vector regression \cite{hahm2016parkinson}. Two limitations are common to this line of work: it relies on disease-specific supervised training with pathological labels, and uses these features jointly with broad audio features.

Despite their promise for predicting speech impairment severity, EMA recordings remain impractical for routine clinical use due to the time required for data collection and analysis, their invasiveness, and the fact that they perturb the speech production process itself \cite{thompson2026effects, Tienkamp2024ISSP_AAVS_EMA}. Recently, speaker-independent acoustic-to-articulatory inversion (AAI) has made it possible to recover EMA-style trajectories from audio alone, which we call quasi-EMA, reaching held-out Pearson correlations of $r{=}0.78$~\cite{wu23_icassp} to $r{=}0.85$~\cite{chung24_interspeech} on the HPRC benchmark. This opens the door to a different scoring approach: rather than training a disease-specific classifier with joint EMA and audio features, the per-frame articulatory trajectory, parameterized as vocal-tract constriction variables (tract variables, TVs), can be compared directly against a typical reference, in the same way that NAD compares self-supervised learning (SSL) features.

In this work, we extend a previous systematic comparison of speech assessment tools \cite{halpern2026pathbench} by showing that \underline{without any pathological training data}, articulatory inversion can provide an interpretable way to measure speech intelligibility. To be precise, our contributions are the following falsifiable claims, each one mapped to a specific row of the results table and a specific paragraph in the discussion:

\noindent\textbf{C1: ART-NAD substantially outperforms the classical vowel space area baseline.} Across 20 protocols, ART-NAD-FA reaches an average $r=0.71$ versus $0.15$ for Vowel Space Area, supporting the use of acoustic-to-articulatory inversion in place of formant-based measures.

\noindent\textbf{C2: ART-NAD reaches the same average correlation as the strongest reference-audio metric, and surpasses it on several protocols.} ART-NAD-FA reaches the same average $r$ as NAD-FA ($0.71$ vs $0.71$) with no significant per-protocol difference (Wilcoxon $p=0.18$), without any pathological-speech training, and is the strongest reference-audio metric on 6 of the 20 protocols. Because both metrics share the same \texttt{wav2vec2-L10} backbone, the average tie isolates the effect of the nine-dimensional TV bottleneck rather than a difference in SSL encoder.

\noindent\textbf{C3: Tract-variable features expose per-constriction, per-phoneme deviations that SSL features cannot.} We illustrate this with a worked example on a UASpeech speaker where the metric localizes the deviation to the tongue-tip and tongue-mid constriction during the word onset.

\section{Methods}

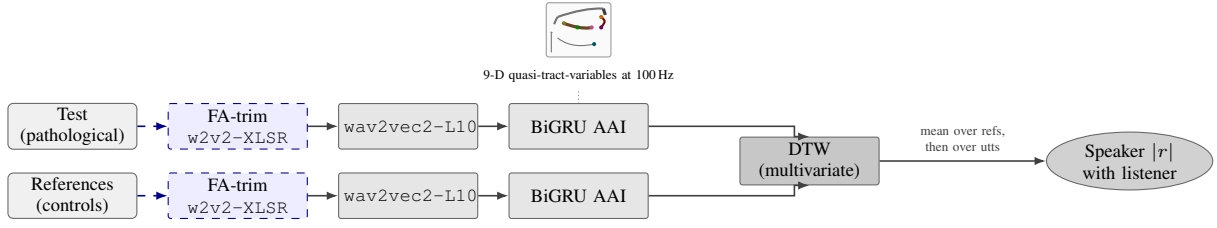
\begin{figure*}[t]
\centering
\begin{tikzpicture}[
    font=\scriptsize,
    >={Latex[length=1.5mm]},
    node distance=3mm and 4mm,
    every node/.style={align=center},
    audio/.style={
        rectangle, rounded corners=2pt, draw=black!60, fill=black!6,
        inner sep=1.5pt, minimum height=5mm, text width=16mm
    },
    feat/.style={
        rectangle, rounded corners=1pt, draw=black!60, fill=black!10,
        inner sep=2pt, minimum height=6mm, text width=17mm
    },
    score/.style={
        rectangle, rounded corners=1pt, draw=black!60, fill=black!22,
        inner sep=2pt, minimum height=5mm, text width=17mm
    },
    optmodel/.style={
        rectangle, rounded corners=1pt, draw=blue!55!black, dashed, fill=blue!7,
        inner sep=2pt, minimum height=6mm, text width=17mm
    },
    aug/.style={
        rectangle, rounded corners=1pt, draw=blue!55!black, dashed, fill=blue!7,
        inner sep=1.5pt, minimum height=4mm, text width=20mm
    },
    outnode/.style={
        ellipse, draw=black!60, fill=black!18,
        inner sep=1pt, minimum height=5mm, minimum width=22mm
    },
    arr/.style={->, semithick, black!75},
    arrdash/.style={->, semithick, blue!55!black, dashed},
]

\node[audio] (test) {Test\\(pathological)};
\node[audio, below=of test] (ref) {References\\(controls)};

\node[optmodel, right=of test] (trimT) {FA-trim\\\texttt{w2v2-XLSR}};
\node[optmodel, right=of ref]  (trimR) {FA-trim\\\texttt{w2v2-XLSR}};

\node[feat, right=of trimT] (hubT) {\texttt{wav2vec2-L10}};
\node[feat, right=of trimR] (hubR) {\texttt{wav2vec2-L10}};

\node[feat, right=of hubT] (aaiT) {BiGRU AAI};
\node[feat, right=of hubR] (aaiR) {BiGRU AAI};

\node[draw=black!40, fill=black!2, rounded corners=1pt, inner sep=1.5pt,
      above=6mm of aaiT] (face) {%
    \begin{tikzpicture}[scale=0.55, baseline=(current bounding box.center)]
        \draw[black!55, thick]
            (1.00, 0.40) .. controls (0.70, 0.42) and (0.30, 0.34) .. (-0.10, 0.20);
        \draw[black!55, thick, dashed]
            (-0.10, 0.20) .. controls (-0.18, 0.05) .. (-0.25, -0.18);
        \draw[black!55] (-0.25, -0.18) -- (-0.25, -0.65);
        \draw[black!75, very thick] (1.00, 0.40) -- (1.05, 0.25);
        \draw[brown!55!black, very thick]
            (0.05, 0.05) .. controls (0.22, -0.02) and (0.55, -0.10) .. (0.72, -0.05);
        \draw[red!45!black, thick]
            (0.95, 0.20) .. controls (1.06, 0.10) .. (0.95, -0.05);
        \draw[black!55]
            (0.78, -0.45) .. controls (0.30, -0.55) and (0.05, -0.42) .. (-0.10, -0.30);
        \fill[teal!70!black]    (0.78, -0.45) circle (1.4pt);
        \fill[orange!80!black]  (0.95,  0.20) circle (1.4pt);
        \fill[violet!80!black]  (0.95, -0.05) circle (1.4pt);
        \fill[magenta!80!black] (0.72, -0.05) circle (1.4pt);
        \fill[green!50!black]   (0.38, -0.05) circle (1.4pt);
        \fill[olive!80!black]   (0.05,  0.05) circle (1.4pt);
    \end{tikzpicture}%
};
\node[font=\tiny, align=center, below=0.4mm of face] (faceCap)
    {9-D quasi-tract-variables at 100\,Hz};
\draw[densely dotted, black!50] (faceCap.south) -- (aaiT.north);

\coordinate (midAAI) at ($(aaiT.east)!0.5!(aaiR.east)$);
\node[score, right=12mm of midAAI] (dtw) {DTW\\(multivariate)};

\node[outnode, right=22mm of dtw] (final) {Speaker $|r|$\\with listener};

\draw[arrdash] (test) -- (trimT);
\draw[arrdash] (ref)  -- (trimR);
\draw[arr] (trimT) -- (hubT);
\draw[arr] (trimR) -- (hubR);
\draw[arr] (hubT)  -- (aaiT);
\draw[arr] (hubR)  -- (aaiR);

\draw[arr] (aaiT.east) -| ([xshift=-2mm]dtw.north) -- (dtw.north);
\draw[arr] (aaiR.east) -| ([xshift=-2mm]dtw.south) -- (dtw.south);

\draw[arr] (dtw) -- node[above, font=\tiny, midway, align=center]
        {mean over refs,\\then over utts}
    (final);

\end{tikzpicture}
\caption{ART-NAD scoring pipeline. Test and same-text reference audios are passed through \texttt{wav2vec2-Large} layer 10 and a BiGRU acoustic-to-articulatory inversion model that emits a 9-channel quasi-TV trajectory at 100\,Hz. The raw channels are aligned by multivariate DTW; distances are averaged across references, then across utterances, giving a speaker-level score correlated with listener intelligibility. The dashed branch marks optional forced-alignment silence trimming. The NAD baseline uses the \texttt{wav2vec2-Large} layer-10 features directly, without the BiGRU.}
\label{fig:pipeline}
\end{figure*}

The full scoring pipeline is shown in Fig.~\ref{fig:pipeline}.

\subsection{Articulatory inversion model}
\label{sec:aai}

We build on the speaker-independent acoustic-to-articulatory inversion model of Wu et al.~\cite{wu23_icassp}. An SSL encoder's hidden state (50~Hz, 1024-D) is linearly interpolated by a factor of two to 100~Hz, then mapped by a two-layer bidirectional GRU to a 9-channel trajectory of TVs \cite{seneviratne19_interspeech}. The nine channels are lip aperture, lip protrusion, jaw angle, and the constriction location and degree of the tongue tip, tongue mid, and tongue back. Unlike raw midsagittal pellet coordinates, TVs parameterize the vocal tract by the location and degree of each constriction, a speaker-normalized, geometry-relative description that maps directly onto phonological constriction targets.

We use \texttt{wav2vec2-Large} layer 10 (\texttt{wav2vec2-L10}) as the SSL encoder for both inversion models we train, matching the backbone of the NAD baseline so that any difference isolates the TV bottleneck rather than the encoder. The two models share the same BiGRU architecture, backbone, and HPRC training data \cite{tiede2017quantifying}, and differ only in their regression target. The first predicts the nine TVs, derived from the corpus EMA following \cite{seneviratne19_interspeech}, and drives ART-NAD; on held-out HPRC speakers it reaches a per-utterance correlation of $r=0.79$ with the reference TVs. The second predicts the twelve raw midsagittal pellet coordinates of the corpus, the $(x,y)$ positions of the jaw, upper lip, lower lip, tongue tip, tongue body, and tongue back, reaching a per-utterance correlation of $r=0.73$ with the reference coordinates on the same held-out speakers, and is used only to compute the KVSA baseline (Sec.~\ref{sec:eval}). We keep this positional model separate because a convex-hull area requires absolute positional coordinates, which the constriction-variable representation does not provide.

\subsection{Proposed method: ART-NAD}
\label{sec:artnad}

The idea of ART-NAD is to use quasi-TV features instead of SSL features for the dynamic time warping step of NAD. The idea is motivated by two facts. First, the SSL features already predict articulation well by simple linear probing, meaning that this should not cause performance degradation \cite{cho2023evidence}. Second, currently SSL-based NAD reach state-of-the-art $r=0.71$ on the speaker intelligibility level, which is much better than what current vowel space area measures are capable of.

\paragraph{Per-utterance features.}
Each utterance was first resampled to 16~kHz and run through the inversion model to obtain a $(T,9)$ TV trajectory at 100~Hz, which is fed to DTW without further normalization. Because TVs are constriction measures expressed in a vocal-tract-relative frame rather than absolute pellet positions, they are largely free of the speaker rest-position bias that affects predicted EMA coordinates~\cite{wu23_icassp}, and their channels share a common physical scale. Empirically, per-utterance or per-speaker normalization of the TV trajectories did not improve the listener-score correlation, so we keep the raw trajectories.

\paragraph{DTW cost.}
Given a test utterance $T$ and a reference utterance $R$, ART-NAD is computed as the length-normalized multivariate DTW distance over the 9-D TV frames, $d(T,R) = \mathrm{DTW}(T, R)$. The per-frame cost is the Euclidean distance between the 9-D TV frames, and the accumulated cost is divided by the length of the warping path; this is the same DTW implementation as used for NAD in PathBench. For consistency with the corresponding reference-audio NAD metric in PathBench~\cite{halpern2026pathbench}, when multiple references were available for the same text we averaged $d(T,R_k)$ over the same-text controls $R_k$ prescribed by the PathBench protocol (ART-NAD-control family).

\paragraph{Silence trimming (-FA models).}
For every reference-audio measure we additionally evaluated a variant that performs leading and trailing silence trimming before scoring, denoted by the ``-FA'' suffix. Trimming was done with a \texttt{wav2vec2-XLSR-53-espeak-cv-ft} forced aligner, mirroring the PathBench NAD-FA setup. PathBench chose forced alignment over a standard voice activity detector because energy-based VAD misclassifies the breathy and hoarse segments that carry the pathology \cite{awan2009estimating}, whereas the aligner uses the transcription and segments by phoneme content; we return to this choice in Sec.~\ref{sec:trimming}.

We used the aligner only to locate the first and last speech frames, and we reported both the untrimmed and silence-trimmed variants of every ART-NAD configuration.

\subsection{Evaluation protocols and baselines}
\label{sec:eval}

We evaluated on all 18 reference-audio protocols of PathBench~\cite{halpern2026pathbench} that have parallel control audio: UASpeech \cite{kim2008dysarthric} (English, word-level, $n{=}14$ speakers with dysarthria), TORGO \cite{rudzicz2012torgo} (English, utterance and word level, $n{=}8$, speakers with dysarthria), COPAS \cite{van2009dutch} (Flemish-Dutch, various pathologies, with $n{=}11$ to $216$ speakers depending on the protocol), EasyCall \cite{turrisi2021easycall} (Italian, $n{=}30$, dysarthria), NeuroVoz \cite{mendes2024neurovoz} (Spanish, $n{=}50$, Parkinson's). 

We additionally created two new protocols based on the Mandarin Dysarthric Speech Corpus (MDSC) \cite{gao2024enhancing}, motivated by the desire to extend the benchmark to tonal languages where the cues that drive intelligibility may differ. MDSC contains $n{=}21$ speakers with dysarthria secondary to cerebral palsy and hepatolenticular degeneration, together with $25$ control speakers. The recordings include multiple repetitions of each item; for the matched control (MC) subset we kept one instance per repetition. Intelligibility ratings came from five expert annotators using a transcription task, an evaluation protocol effectively identical to that of UASpeech. As the sentence/word distinction is less clear for Mandarin than for the other languages, we labeled both MDSC protocols ``Command''.

Following PathBench's convention, ``MC'' denotes the matched control subset, ``EX'' the extended, ``Full'' is using all available data. The reference pool for each test utterance is the set of same-text control recordings prescribed by the PathBench protocol. All metrics were speaker-aggregated by averaging utterance scores, and we reported the Pearson correlation $r$ between the mean of the utterance-level metrics, and the mean of the listener ratings for each speaker.
The sign of the DTW-based metrics (NAD and the ART-NAD family) was flipped so that higher values consistently mean closer agreement with listener scores across all columns.

We compared against three baselines from PathBench~\cite{halpern2026pathbench}:

\noindent\textbf{ArtP}~\cite{talkar2025development}: a phonetic intelligibility score derived from forced alignment of a phoneme target against the CTC emissions of the multilingual \texttt{wav2vec2-XLSR-53-espeak-cv-ft} phonetic recognizer. The target is obtained by phonemizing the reference transcription to IPA with \texttt{phonemizer} \cite{Bernard2021} using the language-specific \texttt{espeak-ng} backend, and the per-utterance score is the mean forced-alignment confidence over the target phonemes excluding silence phonemes. We reported ArtP as the strongest reference-text-aware metric in PathBench and as an upper-bound reading on what current reference-based assessment can do.

\noindent\textbf{VSA (Vowel Space Area)}: computed with the continuous-speech method of Sandoval et al.~\cite{sandoval2013automatic}, in which $F_1$ and $F_2$ are extracted at every voiced frame using Praat \cite{boersma2005praat}, a Gaussian-mixture filter removes outliers, the remaining points are clustered with KMeans initialized at language-specific canonical vowel formants (Italian \cite{bertinetto2005sound}, Spanish \cite{bradlow1995comparative}, Dutch \cite{adank2004acoustic}, Mandarin \cite{chen2001vowel}), and VSA is the convex-hull area of the resulting cluster centroids. We additionally tuned the formant ceiling with the optimizer of Carignan\footnote{\url{github.com/ChristopherCarignan/formant-optimization}}, but this did not improve the listener-score correlation.

\noindent\textbf{KVSA (Kinematic Vowel Space Area)}: the kinematic analogue of VSA, computed on quasi-EMA rather than formants \cite{whitfield2018examining}. KVSA is computed from the positional inversion model of Sec.~\ref{sec:aai} rather than the constriction-variable model used for ART-NAD, since a convex-hull area requires absolute positional coordinates; the $(x, y)$ trajectories of the tongue tip and tongue body are extracted over the whole utterance. Per speaker, we pooled the tongue-tip and tongue-body (x, y) points across all of the speaker's utterances and reported the convex-hull area of the pooled points as the speaker's KVSA, mirroring VSA's speaker-level pooling (formant frames accumulated across utterances before the hull). KVSA therefore isolates what is gained by replacing formants with quasi-EMA positions while keeping the same static, area-based aggregation.

\noindent\textbf{NAD}: the Neural Acoustic Distance of Bartelds et al.~\cite{bartelds2022neural}, a reference-audio metric that quantifies how close a test utterance sounds to a same-text control by aligning their self-supervised representations under Dynamic Time Warping. Each utterance is encoded by \texttt{wav2vec2-Large} into a $(T,1024)$ trajectory, NAD is then the length-normalized DTW distance between the test and reference trajectories, and per-test scores are averaged over all gender-matched same-text controls. We used \texttt{wav2vec2-L10}, following PathBench~\cite{halpern2026pathbench} for its published NAD numbers.

Every reference-audio metric (NAD, ART-NAD) was reported both with and without the forced-alignment silence-trimming step, denoted by the ``-FA'' suffix. Note that the PathBench NAD is NAD-FA.

\paragraph{Reproducibility.}
All ART-NAD evaluator code, configuration, analysis scripts, and the retrained inversion are available as part of the PathBench repository\footnote{\url{https://github.com/karkirowle/pathbench}}. The inversion architecture of Wu et al.~\cite{wu23_icassp}, the \texttt{wav2vec2-XLSR-53-espeak-cv-ft} forced aligner, and the \texttt{wav2vec2-Large} encoder are all publicly available HuggingFace assets.

\section{Results and discussion}

Table~\ref{tab:art_nad} reports speaker-level $r$ across the 20 protocols for ArtP, the two vowel space area baselines (VSA, KVSA), the two NAD variants, and the two ART-NAD variants (with and without silence trimming). ArtP is reference-text, scoring against the intended transcription, whereas the NAD and ART-NAD families are reference-audio, scoring against control recordings; ArtP therefore has more information and reads as an upper bound rather than a like-for-like comparison.

\begin{table*}[t]
\centering
\caption{Speaker-level Pearson correlation with listener scores ($r$) on the 20 PathBench reference-audio protocols. Sign of DTW-based metrics is flipped for visual parity. Per row, \textbf{bold} marks the highest $r$ across all columns, and \underline{underline} marks the highest $r$ among the reference-audio metrics (NAD, NAD-FA, ART-NAD, ART-NAD-FA). A cell carrying both styles tops both subsets simultaneously.}
\label{tab:art_nad}
\setlength{\tabcolsep}{3pt}
\begin{tabular}{lll l ccccccc}
\toprule
Dataset & Language & Protocol & Type & ArtP & VSA & KVSA & NAD & NAD-FA & ART-NAD & ART-NAD-FA \\
\midrule
\multirow{2}{*}{UASpeech} & \multirow{2}{*}{English} & Word & MC & \textbf{0.98} & 0.02 & 0.80 & 0.90 & \underline{0.97} & 0.90 & 0.92 \\
          &          & Word & EX & \textbf{0.98} & -0.08 & 0.80 & 0.89 & \underline{0.97} & 0.90 & 0.94 \\
\midrule
\multirow{2}{*}{NeuroVoz} & \multirow{2}{*}{Spanish} & Sentence & MC & \textbf{0.77} & 0.01 & 0.40 & 0.64 & \underline{0.70} & 0.54 & 0.62 \\
          &          & Sentence & EX & \textbf{0.78} & 0.05 & 0.35 & 0.71 & \underline{0.75} & 0.54 & 0.62 \\
\midrule
\multirow{4}{*}{EasyCall} & \multirow{4}{*}{Italian} & Word & MC & 0.71 & 0.27 & 0.27 & 0.57 & \textbf{\underline{0.76}} & 0.46 & 0.71 \\
          &          & Word & EX & 0.74 & 0.19 & 0.26 & 0.66 & \textbf{\underline{0.83}} & 0.61 & 0.76 \\
          &          & Sentence & MC & 0.69 & 0.15 & 0.32 & 0.72 & \textbf{\underline{0.78}} & 0.56 & 0.72 \\
          &          & Sentence & EX & 0.68 & 0.28 & 0.24 & 0.79 & \textbf{\underline{0.86}} & 0.62 & 0.71 \\
\midrule
\multirow{6}{*}{COPAS} & \multirow{6}{*}{Dutch} & Word & MC & 0.40 & -0.20 & 0.19 & 0.18 & 0.18 & \textbf{\underline{0.62}} & \textbf{\underline{0.62}} \\
          &          & Word & EX & 0.78 & 0.55 & 0.33 & 0.46 & 0.46 & \textbf{\underline{0.83}} & \textbf{\underline{0.83}} \\
          &          & Word & Full & \textbf{0.61} & 0.04 & 0.18 & 0.51 & 0.51 & \underline{0.58} & \underline{0.58} \\
          &          & Sentence & MC & 0.57 & 0.22 & 0.12 & 0.67 & \textbf{\underline{0.69}} & 0.64 & 0.62 \\
          &          & Sentence & EX & 0.57 & 0.22 & 0.12 & 0.67 & \textbf{\underline{0.69}} & 0.64 & 0.62 \\
          &          & Sentence & Full & 0.56 & 0.20 & 0.12 & 0.67 & \textbf{\underline{0.70}} & 0.63 & 0.61 \\
\midrule
\multirow{4}{*}{TORGO} & \multirow{4}{*}{English} & Word & MC & 0.56 & \textbf{0.63} & 0.48 & 0.40 & 0.52 & 0.49 & \underline{0.56} \\
          &          & Word & EX & 0.61 & 0.58 & 0.48 & 0.63 & 0.55 & \textbf{\underline{0.68}} & 0.56 \\
          &          & Sentence & MC & 0.92 & -0.45 & 0.86 & 0.90 & 0.90 & \textbf{\underline{0.97}} & 0.94 \\
          &          & Sentence & EX & 0.92 & -0.38 & 0.79 & 0.89 & 0.90 & \textbf{\underline{0.95}} & \textbf{\underline{0.95}} \\
\midrule
\multirow{2}{*}{MDSC} & \multirow{2}{*}{Mandarin} & Command & MC & \textbf{0.75} & 0.39 & 0.29 & 0.63 & \underline{0.71} & 0.55 & 0.63 \\
          &          & Command & EX & \textbf{0.73} & 0.38 & 0.30 & 0.60 & 0.71 & 0.67 & \underline{0.72} \\
\midrule
\multicolumn{4}{l}{Average} & \textbf{0.72} & 0.15 & 0.38 & 0.65 & \underline{0.71} & 0.67 & \underline{0.71} \\
\bottomrule
\end{tabular}
\end{table*}

\subsection{C1: ART-NAD outperforms classical vowel space area baselines}

The VSA baseline averages $r=0.15$ across the 20 protocols, with only TORGO Word MC ($0.63$) clearing $r>0.6$, and is negatively correlated with intelligibility on TORGO Sentence ($r=-0.45$ MC, $r=-0.38$ EX) and COPAS Word MC ($r=-0.20$). Routing the same area-summary idea through the AAI model lifts the baseline substantially: KVSA averages $0.38$, more than doubling VSA, and reaches $r=0.80$ on UASpeech Word, $0.86$ on TORGO Sentence MC and $0.79$ on TORGO Sentence EX -- protocols where the formant-based VSA is near zero or negative. The gain is largest where the AAI model was trained (English, both word and sentence) and on NeuroVoz and EasyCall, where the predicted quasi-EMA range tracks intelligibility despite a less canonical formant geometry; KVSA only loses to VSA on COPAS Word EX/Full ($0.33$/$0.18$ vs $0.55$/$0.04$) and TORGO Word ($0.48$ vs $0.63$/$0.58$). ART-NAD-FA in particular exceeds VSA by $0.56$ and KVSA by $0.33$ on average. It appears that the AAI step alone, even reduced to a static convex-hull area, already captures a substantial slice of the articulation that formant geometry misses, and the per-frame, per-constriction alignment against the reference in ART-NAD adds another step on top. The advantage of ART-NAD-FA holds across protocols against both VSA ($18/20$ protocols) and KVSA ($20/20$ protocols), Wilcoxon signed-rank $p<0.001$ for each; the KVSA-over-VSA gain, by contrast, is only an average effect and is not significant per protocol ($p=0.12$, $10/20$), as KVSA improves on VSA mainly where formant geometry fails. This confirms C1.

\subsection{C2: ART-NAD matches NAD-FA on the same backbone}

ART-NAD-FA reaches $r=0.71$ on average, equal to NAD-FA ($0.71$), without any pathological-speech training and on the same \texttt{wav2vec2-L10} backbone that NAD uses. Across the 20 protocols, the per-protocol difference is not statistically significant (Wilcoxon signed-rank $p=0.18$). ART-NAD-FA matches or exceeds NAD-FA on UASpeech (within $0.05$), TORGO, MDSC Command EX, and especially COPAS Word; it lags on NeuroVoz, EasyCall, and COPAS Sentence, where the \texttt{wav2vec2} features capture acoustic-channel cues that the TV bottleneck discards. On 6 of the 20 protocols ART-NAD-FA is the strongest reference-audio metric (all three COPAS Word, TORGO Word MC, TORGO Sentence EX, and MDSC Command EX), and on three of these (COPAS Word MC, COPAS Word EX, and TORGO Sentence EX) it is the top-scoring metric overall, beating even the reference-text ArtP upper bound on the two COPAS Word protocols.

Because the inversion model and NAD share the \texttt{wav2vec2-L10} backbone, the average tie isolates the effect of the TV bottleneck itself rather than a difference in SSL encoder. The bottleneck is not uniformly costly. On COPAS Word, where same-text controls are short and acoustically variable, ART-NAD-FA ($0.62$, $0.83$, $0.58$ for MC, EX, Full) far exceeds NAD-FA ($0.18$, $0.46$, $0.51$), the TV representation being more robust than raw SSL-DTW on short items. This closes C2.

\subsection{C3: Per-articulator, per-phoneme interpretability}

\begin{figure}[t]
\centering
\includegraphics[width=\columnwidth]{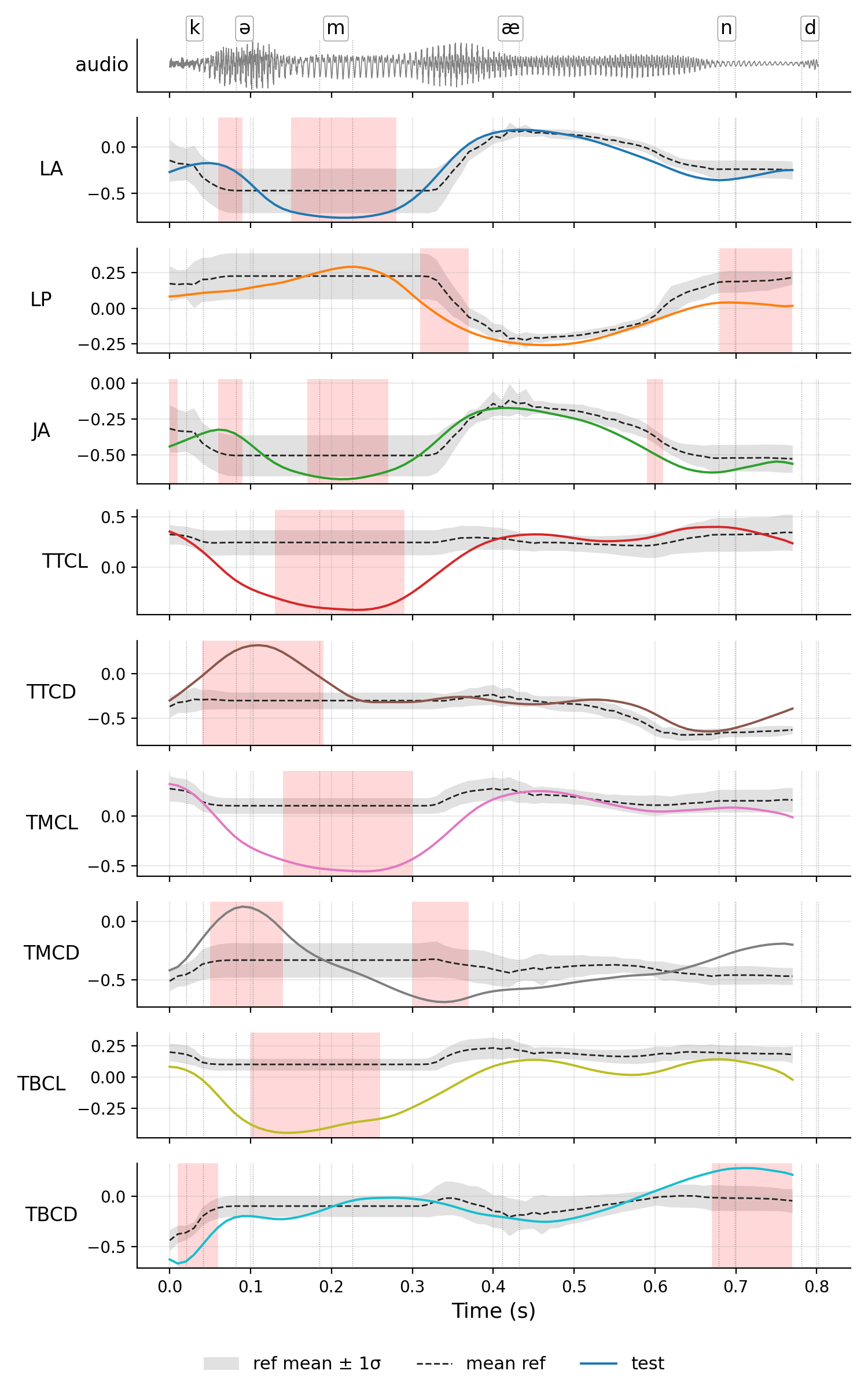}
\caption{ART-NAD interpretability example. UASpeech speaker M07 (cerebral palsy, intelligibility 28/100) saying ``command'' against the mean of nine male control speakers. Each row plots one raw TV: test (colored), DTW-aligned mean reference (dashed), and across-reference $\pm 1\sigma$ band (gray). The nine rows are lip aperture (LA), lip protrusion (LP), jaw angle (JA), and the constriction location and degree of the tongue tip (TTCL, TTCD), tongue mid (TMCL, TMCD), and tongue back (TBCL, TBCD). Red marks frames where the residual between test and mean reference exceeds that channel's 80th percentile over the utterance. Forced-alignment phoneme intervals (k, \textipa{@}, m, {\ae}, n, d) overlay the waveform. The largest deviations fall on the tongue-tip and tongue-mid constriction through the /k-\textipa{@}-m/ onset, indicating imprecise lingual constriction of the /k/, as well as a more posterior realisation of /\textipa{@}/ due to reduced tongue mobility.}
\label{fig:art_nad_example}
\end{figure}

Fig.~\ref{fig:art_nad_example} illustrates the practical benefit of routing the metric through articulation. For every test utterance, ART-NAD compares a per-frame, per-constriction signal against the aligned mean reference. On M07's (cerebral palsy, intelligibility 28/100) ``command,'' the red bands concentrate on the lingual constriction variables through the initial /k-\textipa{@}-m/ stretch: the tongue-tip and tongue-mid constriction location and degree depart from the control band while the references hold a stable constriction, and the lip aperture and jaw angle deviate alongside. Because each channel is a named constriction rather than an opaque feature, and the forced-aligner phoneme intervals are overlaid, the clinician can read off a statement such as ``the tongue-tip and tongue-mid constriction is imprecise through the /k-a-m/ onset of the word command,'' which points to the articulatory target that was produced more posteriorly due to reduced tongue mobility and limited differentiation between tongue tip, dorsum, and dorsum movements, rather than merely a time range. The same attribution is not available for NAD: \texttt{wav2vec2} features have no articulatory decomposition, so NAD can highlight a frame range as problematic but cannot attribute the problem to a specific constriction. TVs also read more directly than raw predicted EMA coordinates: a constriction degree answers a clinically direct question (was the closure achieved?), whereas a pellet $y$-coordinate must first be related to a target position. This articulatory attribution is what makes TV features valuable for clinical adoption even when the headline correlation does not exceed the SSL baseline, closing C3. We show a single illustrative example here; a quantitative validation of the localized attributions against clinical judgment is left to future work.

\subsection{Effect of language}

The per-protocol pattern in Table~\ref{tab:art_nad} tracks the inversion model's training language, but most clearly on connected speech. On the sentence protocols ART-NAD-FA matches or exceeds NAD-FA only on English (TORGO Sentence, $+0.04$ and $+0.05$, despite a $9$-D TV bottleneck against NAD's $1024$-D \texttt{wav2vec2} features), and trails on every language the inversion model has not seen: by about $0.08$--$0.13$ on Spanish (NeuroVoz), $0.07$ on Flemish-Dutch (COPAS Sentence), $0.06$--$0.15$ on Italian (EasyCall Sentence), and $0.09$ on Mandarin (MDSC Command MC). The Dutch drop is consistent with Hao et al., who report a significant inversion degradation for Dutch when training on English \cite{hao2024exploring}; measuring that drop directly would need ground-truth EMA. The isolated-word protocols are governed less by language than by item length and corpus: ART-NAD-FA beats NAD-FA by up to $0.44$ on the COPAS Word protocols, an unseen language, where short and acoustically variable controls favor the TV representation, and edges ahead on the English TORGO Word, yet trails slightly on UASpeech Word. On Mandarin MDSC, the most distant language, the inversion-based scores stay below the English protocols across all metrics.


\subsection{Effect of silence trimming}
\label{sec:trimming}

Whether to apply silence trimming is a difficult question. Reference-audio metrics like NAD and ART-NAD score every frame, so silent regions that appear in both test and reference align trivially under DTW and inflate the apparent similarity, suppressing the true intelligibility signal. At the same time, confounders such as microphone quality and recording conditions can make pathology partly detectable from silence segments alone, an effect documented for TORGO and UASpeech in particular \cite{schu2023using}, so deciding what to keep and what to drop is itself a modeling choice. No existing voice activity detection (VAD) method annotates silence reliably for pathological speech: energy-based VAD is unreliable for dysphonic speech \cite{awan2009estimating}, neural VAD trained on typical speech is unlikely to generalize, and manual VAD is hard to trust without an inter-rater reliability estimate. ASR-based forced alignment is the alternative we prefer, as the aligner's sensitivity usually gets worse with speaker severity, which is in itself a useful signal for intelligibility estimation.

Empirically, trimming helps both reference-audio metrics by a similar amount. Comparing each metric against its own trimmed variant across the 20 protocols with a paired Wilcoxon signed-rank test, NAD improves from $0.65$ to $0.71$ ($p=0.003$) and ART-NAD from $0.67$ to $0.71$ ($p=0.03$); the per-metric trimming gain is significant in both cases. Removing leading and trailing silence, where test and reference align trivially under DTW, sharpens the distance for both feature types, and it is this trimming step that brings ART-NAD level with NAD-FA.

The case for trimming is also measure-dependent. Pre-initiation segments such as effort breaths are irrelevant when the listener judges the phonetic content of a target word, but they carry information for measures such as speaker typicality, naturalness, or initiation cost. The FA-trim variant of ART-NAD should therefore be read specifically as an intelligibility-scoring variant, and we leave a systematic study of VAD strategies for pathological speech to future work.

\subsection{Limitations and future work}

The inversion model is trained on English EMA only, which as we have seen causes performance degradation for non-English languages. We plan to incorporate the EMA data from Dutch speakers treated for oral cancer \cite{halpern2022manipulation} in future work. A second practical limitation is that reference-audio results are sensitive to the low number of control speakers available in several PathBench protocols, which could be alleviated in the future by appropriate text-to-speech methods to synthesize additional matched controls. Finally, this work inherits the limitations of PathBench itself, in particular the assumption that the different listener-elicited percepts in the collection are largely equivalent \cite{halpern25_interspeech}. The fact that we obtain moderate-to-strong correlations across protocols that differ in the percept being rated suggests that this assumption is less restrictive in practice than it might seem.


\section{Conclusion}

We introduced ART-NAD, a reference-audio intelligibility metric that aligns quasi-TV trajectories from a speaker-independent acoustic-to-articulatory inversion model. Without any pathological-speech training, ART-NAD-FA reaches an average $r=0.71$ across 20 PathBench reference-audio protocols, equal to NAD-FA ($0.71$; Wilcoxon $p=0.18$) on the same \texttt{wav2vec2-L10} backbone and significantly above the formant-based VSA baseline ($0.15$; $p<0.001$). It thus matches the strongest reference-audio metric at no accuracy cost while adding interpretability; the main remaining limitation is language, where the gap to NAD-FA widens on languages the inversion model has never seen. Apart from the score itself, the metric exposes which vocal-tract constriction deviates from the reference at specific phoneme demarcated time points, providing a clinician-readable explanation that no SSL-feature reference-audio metric currently offers.

\section*{Acknowledgment}
This work is partly financed by the Dutch Research Council (NWO) under project number 019.232SG.011 and 019.252SG.013. This work was also supported in part by JSPS KAKENHI under Grant 26H02530, and in part by the BRIDGE Program (R7-H05), implemented by the Cabinet Office, Government of Japan.
During the preparation of this work the authors used Claude (Anthropic) for prose editing and supporting analysis scripts. All generated content was reviewed by the authors.

\bibliographystyle{IEEEtran}
\bibliography{literature}

\end{document}